\documentclass[aps,twocolumn,amsmath,amssymb,groupedaddress,pre,floatfix,10pt]{revtex4-1}

\usepackage{graphicx} 
\usepackage{subfigure}
\usepackage[dvipsnames]{xcolor}
\usepackage{wasysym}
\usepackage{dcolumn}
\usepackage{epstopdf}
\usepackage{hyperref}
\usepackage{textcomp} 
\usepackage{gensymb}
\usepackage{soul}

\hypersetup{
    colorlinks=true,       
    linkcolor=blue,          
    citecolor=blue,        
    filecolor=blue,      
    urlcolor=blue           
}

\begin{document}

\title{Plasticity and Aging of Folded Elastic Sheets}

\author{T. Jules\textsuperscript{1,2}}
\email{theo.jules@ens-lyon.fr}
\author{F. Lechenault\textsuperscript{2}}
\author{M. Adda-Bedia\textsuperscript{1}}

\affiliation{\textsuperscript{1}Universit\'e de Lyon, Ecole Normale Sup\'erieure de Lyon, Universit\'e Claude Bernard, CNRS, Laboratoire de Physique, F-69342 Lyon, France}
\affiliation{\textsuperscript{2}Laboratoire de Physique de l'\'Ecole Normale Sup\'erieure, ENS, PSL Research University, CNRS, Sorbonne University, Université Paris Diderot, Sorbonne Paris Cité, 75005 Paris, France}

\date{\today}

\begin{abstract}
We investigate the dissipative mechanisms exhibited by creased material sheets when subjected to mechanical loading, which comes in the form of plasticity and relaxation phenomena within the creases. After demonstrating that plasticity mostly affects the rest angle of the creases, we devise a mapping between this quantity and the macroscopic state of the system that allows us to track its reference configuration along an arbitrary loading path, resulting in a powerful monitoring and design tool for crease-based metamaterials. Furthermore, we show that complex relaxation phenomena, in particular memory effects, can give rise to a non-monotonic response at the crease level, possibly relating to the similar behavior reported for crumpled sheets. We describe our observations through a classical double-logarithmic time evolution and obtain a constitutive behavior compatible with that of the underlying material. Thus the lever effect provided by the crease allows magnified access to the material's rheology.
\end{abstract}

\maketitle

Systematically creasing a thin material sheet can produce a variety of bulk metamaterials, that can be naturally divided into disordered, or {\it crumpled}, and ordered, {\it origami-like}, structures.
On the origami side, carefully picking among the infinite number of possible crease patterns allows designing a wide range of physical properties and shapes~\cite{Dudte2016, Overvelde2016, Silverberg2014, Filipov2015, Dias2012}. An archetypal example is the negative apparent Poisson ratio exhibited by the Miura-ori patterns~\cite{Schenk2013, Lv2014, Wei2013, Yasuda2015}. This unnatural behavior is explained through the \emph{rigid-face} model, where each fold is described as two rigid panels and a hinge setting an angle between them. This description results in tight kinetic constraints on extended foldings, and only a small number of degrees of freedom usually account for all possible geometric deformations~\cite{Schenk2013}. This simple model is functional regardless of the scale of the system, from micro-robots~\cite{Miskin2018} to space engineering~\cite{Miura1985}. In crumpled systems, the situation is more complex, as single elastic excitations such as developable cones and ridges~\cite{Cerda_1999, Lobkovsky_1995} act as crease precursors. However, once the system has been prepared, a random network of crease-like, plastified objects competes with the elasticity of the sheet to produce a soft elastic solid, though in this case, self-contact plays a major role~\cite{Vliegenthart_2006,Deboeuf2013}.

Thus, for many material foldings, the modeling must take into account the properties of the material itself~\cite{Habibi2017}. Indeed, for origami structures hidden degrees of freedom~\cite{Silverberg2015, Grey2019} appear that combine the elastic deformation of the faces and the mechanical response of the creases. The former is understood as a classic deformation of thin sheets with boundary conditions imposed by the creases. The latter is usually described as an elastic hinge, with a response that is proportional to the departure of the crease angle from a rest configuration~\cite{Brunck2016, Waitukaitis2015}. The range of reachable configurations for such a model is much broader: it allows, for instance, the passage between stable states in bistable origamis~\cite{Reid2017, AndradeSilva2019, Walker2018}. Notably, comparing the elasticity of the crease and that of the faces gives rise to a characteristic length-scale~\cite{Lechenault2014} that relates to the spatial extension of the crease~\cite{Jules2019, Walker2019}.

While this approach is enough to explain the elastic behavior of folded structures, it fails to capture the complete quantitative picture. For intermediate deformations, both origamis~\cite{Reid2017, Thiria2011, Farain2020} and crumpled sheets~\cite{Matan2002, Lahini2017} exhibit hysteresis and relaxation. These phenomena drastically limit the experimental domain of validity for simple elastic models: they induce a temporal evolution of the system and a change of reference state during the experiment. Worse, producing precise and reproducible experiments is severely challenging due to induced memory effects. Nevertheless, the ability to produce a crease within a sheet relies on these very effects~\cite{Dharmadasa2020, Farain2020, Benusiglio2012}, which are, in turn, unavoidable. It is thus of crucial interest to disentangle the respective roles of elasticity and dissipative phenomena to understand the macroscopic mechanical behavior of real-world foldings.

In this paper, we first extend the purely elastic description of a single fold to take into account the plasticity of the material through the modification of its reference state. This approach produces a mapping of the load-deformation curve to the rest angle of the crease, allowing to read the latter from macroscopic observations on the fly. The corresponding predictions are then compared to experimental measurements of a single polymeric fold with remarkable success. Finally, we investigate the temporal evolution of the system under stress. A constant macroscopic strain imposes a stress relaxation that is well described by a double logarithm. Such a description, based on the aging of glassy polymers~\cite{Hutchinson1995, Amir2012} and already observed in various complex systems~\cite{Bruggen2019, Lahini2017}, allows capturing subsequent memory effects under complex loading paths, and their dependence upon various parameters.

\begin{figure}[htb]
\begin{center}
\includegraphics[width=0.9\linewidth]{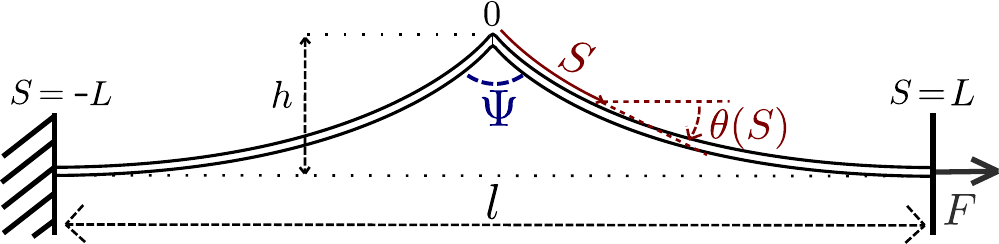}
\caption{Schematics of the experimental setup used to probe the mechanical response $F(l)$ of a creased sheet. One end of the fold is clamped to a rigid wall, and the other one is fixed to a loading device. The instantaneous shape $\theta(S)$ along the curvilinear coordinate $S$ is recovered from direct imaging of the fold~\cite{Lechenault2014,Jules2019}. Notice the mirror symmetry of the system with respect to $S=0$.}
\label{fig:SchemaMontage}
\end{center}
\end{figure}

\section{Mapping the intrinsic parameters of a crease}
\label{sec:Mapping}

The experimental system displayed in Fig.~\ref{fig:SchemaMontage} is similar to the one used in~\cite{Lechenault2014,Jules2019}. Consider a thin sheet of thickness $e$ and size $2L\times W$ decorated with a single crease across its width $W$. The folded sample is clamped at both ends located at an imposed spacing of $l\leq 2L$ while the corresponding external load $F(l)$ is recorded (or vice-versa).
The absolute reference configuration of the free-standing fold prior to the mechanical testing is described by a tent-like shape $\theta_0(S)$ that is well parametrized by
\begin{align}
\theta_0(S) = \cfrac{\Psi_0-\pi}{2}\,\text{tanh}\left(\cfrac{S}{S_0}\right), \label{eqn:ref_state}
\end{align}
with $\Psi_0$ the crease rest angle and $S_0$ its characteristic size. The latter two intrinsic parameters determine the internal state of the fold unambiguously. Recall that the reference configuration of the clamped crease is different from the free-standing one as it is characterized by a size $S_c$ given by
\begin{align}
S_c\simeq\frac{S_0}{2}\log\frac{4L}{S_0}.
\end{align}
and a crease rest angle $\Psi_c=\psi(S_c)\simeq\Psi_0$~\cite{Jules2019}. In the deformed configuration, we assume a local elastic energy density of the fold given by~\cite{Jules2019}
\begin{align}
E_{loc}(S) = \cfrac{BW}{2}\left[\theta'(S)-\theta_0'(S)\right]^2\label{eqn:loc_energy},
\end{align}
where primes denote derivative with respect to $S$ and $B$ is the bending stiffness of the unfolded sheet. The equilibrium configuration of the fold is retrieved by minimizing its total energy elastic energy minus the work of the external force $F$. The minimization yields the following pre-strained elastica~\cite{Jules2019}
\begin{align}
\theta''(s)-\theta_0''(s) - \alpha\sin \theta(s) = 0, \label{eqn:PSE}
\end{align}
with $\alpha=\cfrac{FL^2}{BW}$ the dimensionless load and $s = \cfrac{S}{L}$ the dimensionless curvilinear coordinate. In addition, the boundary conditions at the clamped edges impose
\begin{align}
\theta(0) = \theta(1) = 0, \label{eqn:BC}
\end{align}
where the mirror symmetry of the fold at $s=0$ was used (see Fig.~\ref{fig:SchemaMontage}). The differential equation Eq.~(\ref{eqn:PSE}) with boundary conditions~(\ref{eqn:BC}) can be solved numerically using a standard shooting method. Solving for a prescribed range of values of $\alpha$ yields a load-displacement curve for the clamped fold. To this purpose, we define the typical deformations in the parallel and normal directions to the applied load by
\begin{eqnarray}
\epsilon_x = \frac{l}{2L} &=& \int_0^1 \cos\theta(s)\,ds, \label{eqn:l}\\
\epsilon_y = \frac{h}{L} &=& -\int_0^1 \sin\theta(s)\,ds. \label{eqn:h}
\end{eqnarray}
Here, $\epsilon_x^{(c)}(\Psi_0)<\epsilon_x<1$ and $0<\epsilon_y<\epsilon_{y}^{(c)}(\Psi_0)$, where $\epsilon_x^{(c)}$ and $\epsilon_{y}^{(c)}$ are limiting strains for which self-contact between the two faces of the fold occurs.

\begin{figure}[htb]
\begin{center}
\includegraphics[width=0.9\linewidth]{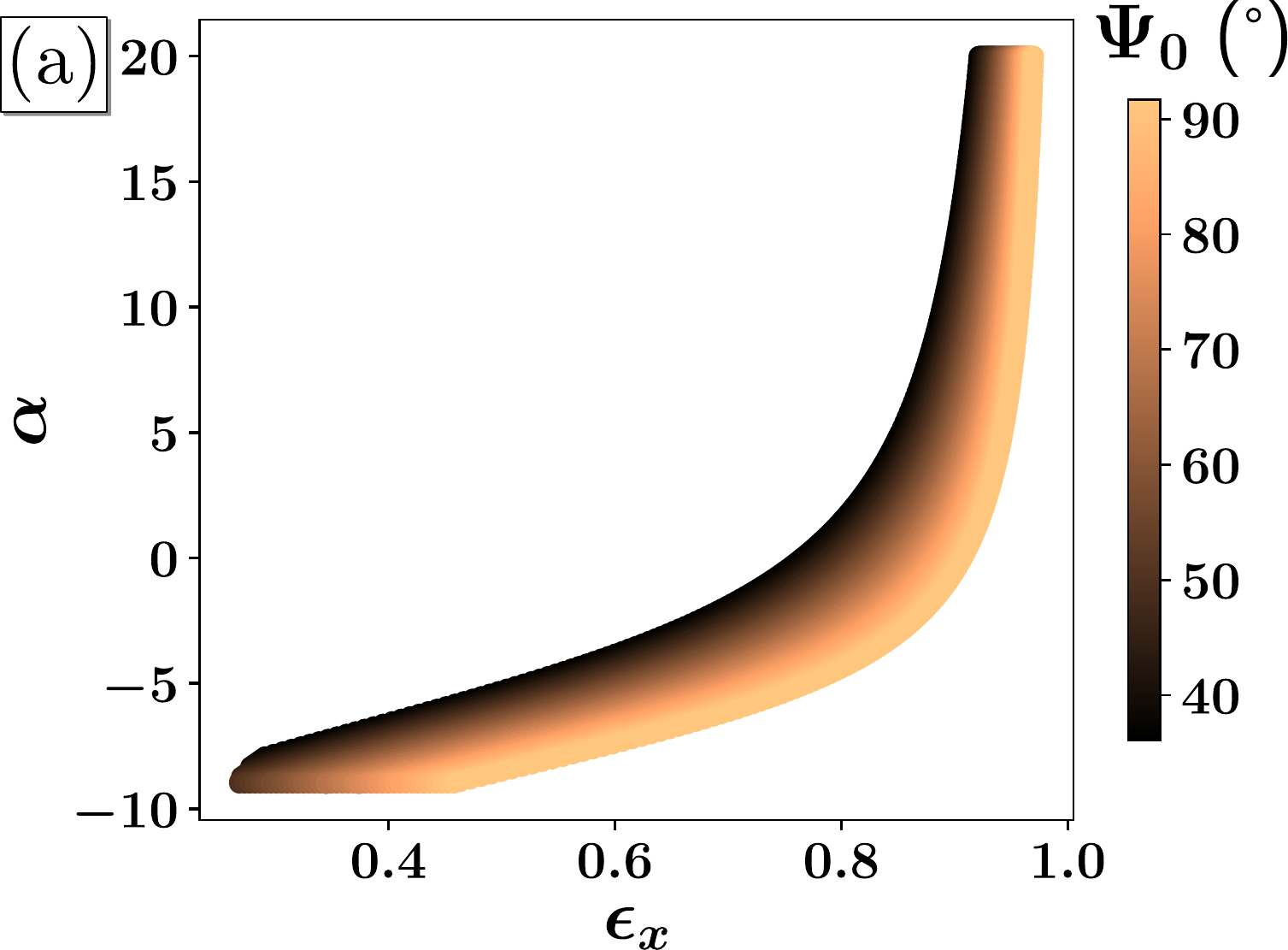}
\includegraphics[width=0.8\linewidth]{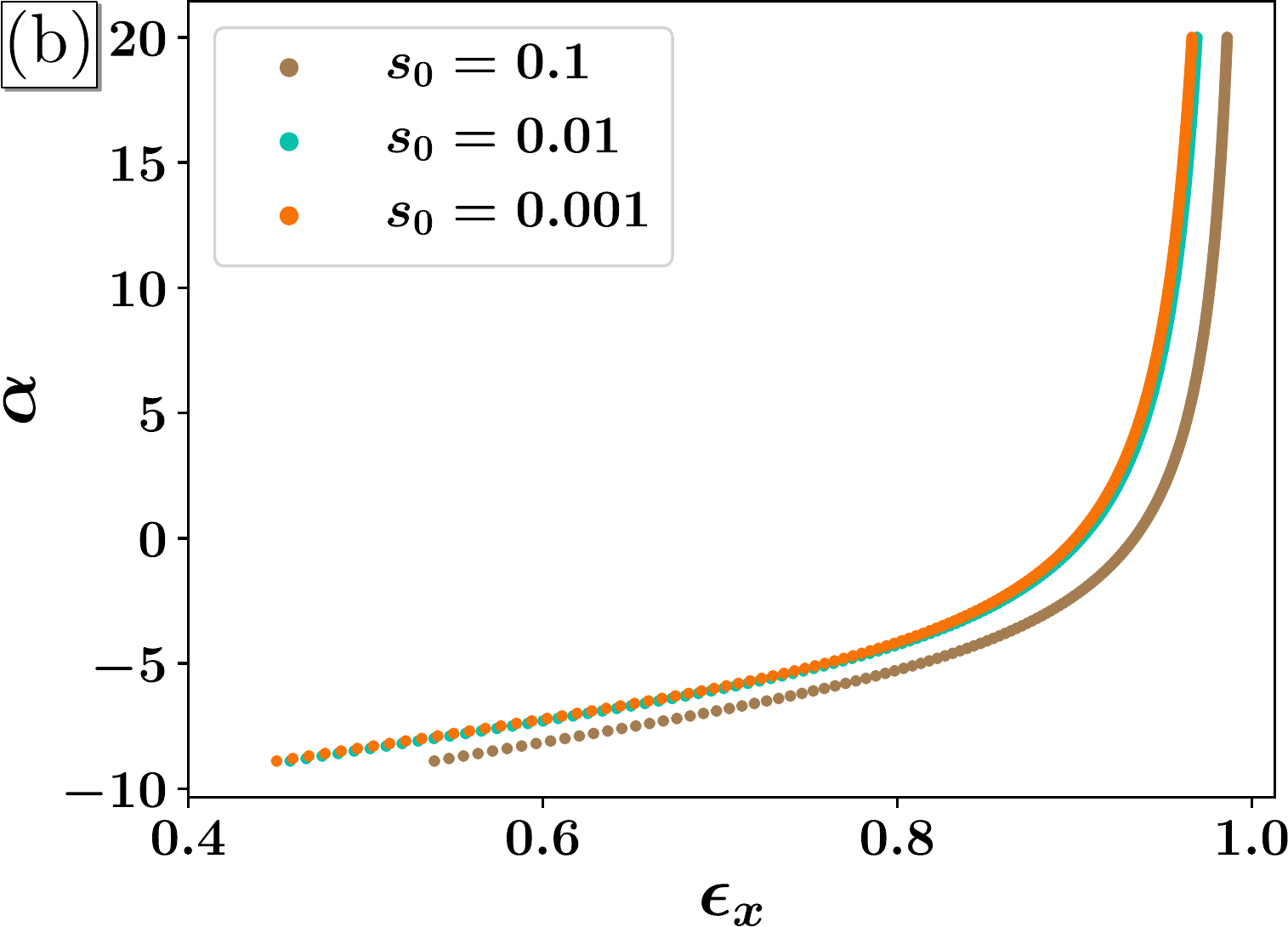}
\caption{(a) Mapping of the load-deformation curve for folds with intrinsic parameters $s_0=9.2 \times 10^{-3}$ and $\Psi_0$ ranging from 36\textdegree\,to 92\textdegree. (b) Mapping of the load-displacement curve for folds with intrinsic rest angle $\Psi_0 = 90$\textdegree\, and three different crease sizes. Notice that these mappings exclude unphysical self-intersecting state of the fold, that is the condition $\epsilon_x^{(c)}(\Psi_0)<\epsilon_x<1$ is always satisfied.}
\label{fig:Mapping}
\end{center}
\end{figure}

Eq.~(\ref{eqn:ref_state}) shows that the absolute reference state is described using two internal parameters $\Psi_0$ and $s_0 = S_0/L$ only. Therefore, one can systematically map their respective effects on the load-displacement curve, as shown in Fig.~\ref{fig:Mapping}. We notice that for a fixed value of $s_0$, each couple ($\alpha$, $\epsilon_x$) corresponds to a unique value of $\Psi_0$: the rest angle is thus traceable through the values of the load $\alpha$ and the corresponding deformation $\epsilon_x$. However, a degeneracy still exists if one considers both parameters $\Psi_0$ and $s_0$.  Nevertheless, there exists a separation of scales between variations due to each parameter:  Fig.~\ref{fig:Mapping}(b) shows that a small shift in the curve of the mechanical response necessitates an order of magnitude variation in $s_0$ (as long as $s_0\ll1$). In contrast, a small variation of the angle $\Psi_0$ induces a substantially larger effect. In the following, we build on this feature to lift this degeneracy by using a direct experimental estimation of the characteristic size $S_0$ and determining $\Psi_0$ from the force-deformation curve.

\begin{figure}[htb]
\begin{center}
\includegraphics[width=0.9\linewidth]{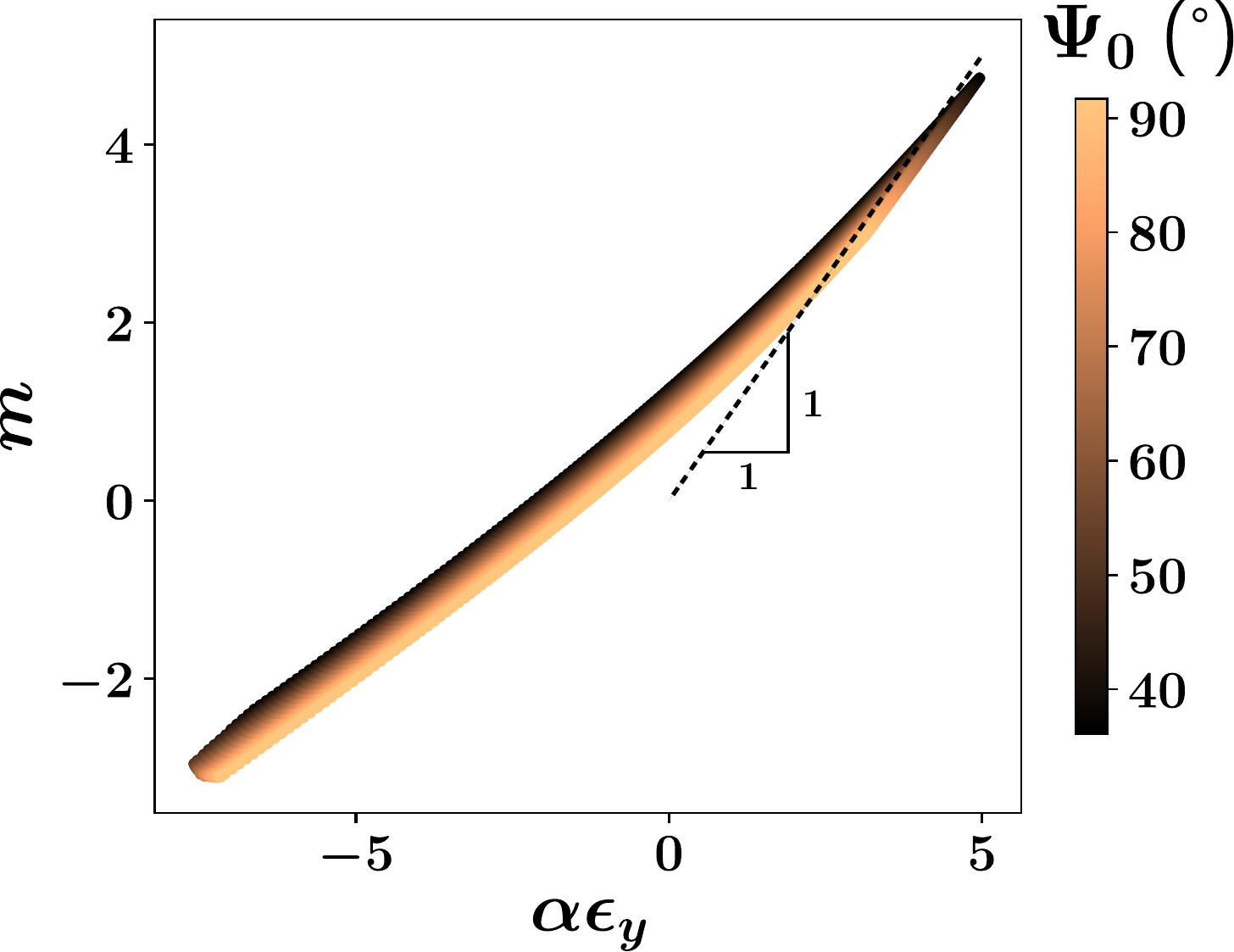}
\caption{Mapping of the normalized crease moment with respect to the macroscopic moment imposed on a fold with a crease size $s_0=9.2 \times 10^{-3}$ and a rest angle $\Psi_0$ ranging from 36\textdegree\,to 92\textdegree. For high stretching, the moment of the crease is close to the one estimated macroscopically.}
\label{fig:moment_micro_macro}
\end{center}
\end{figure}

Interestingly, the computation also allows us to access the dimensionless moment of the crease~\cite{Jules2019}
\begin{equation}
m = \theta'(s_c)-\theta'_0(s_c)\simeq \frac{1}{2s_c}(\Psi-\Psi_0).
\end{equation}
Here, $s_c=S_c/L$, $\Psi=\pi+2\theta(s_c)$ and $BW/2L$ is used as the scaling factor of the moment. In Fig.~\ref{fig:moment_micro_macro}, we compare the moment $m$ to a macroscopic moment, evaluated by multiplying the dimensionless load to the normalized height of the fold, and used as an approximation for the crease mechanical response in the case of highly stretched creases in~\cite{Lechenault2014}. Our model not only confirms that $\alpha\epsilon_y$ is a good approximation of the moment at the crease for high strain but also takes into account both the correction coming from the spatial extension of the crease and the bending of the faces for low and negative loads. We argue that the mapping procedure proposed here from the direct results of the loading test is well suited to characterize the crease mechanics.

\begin{figure}[htb]
\begin{center}
\includegraphics[width=0.8\linewidth]{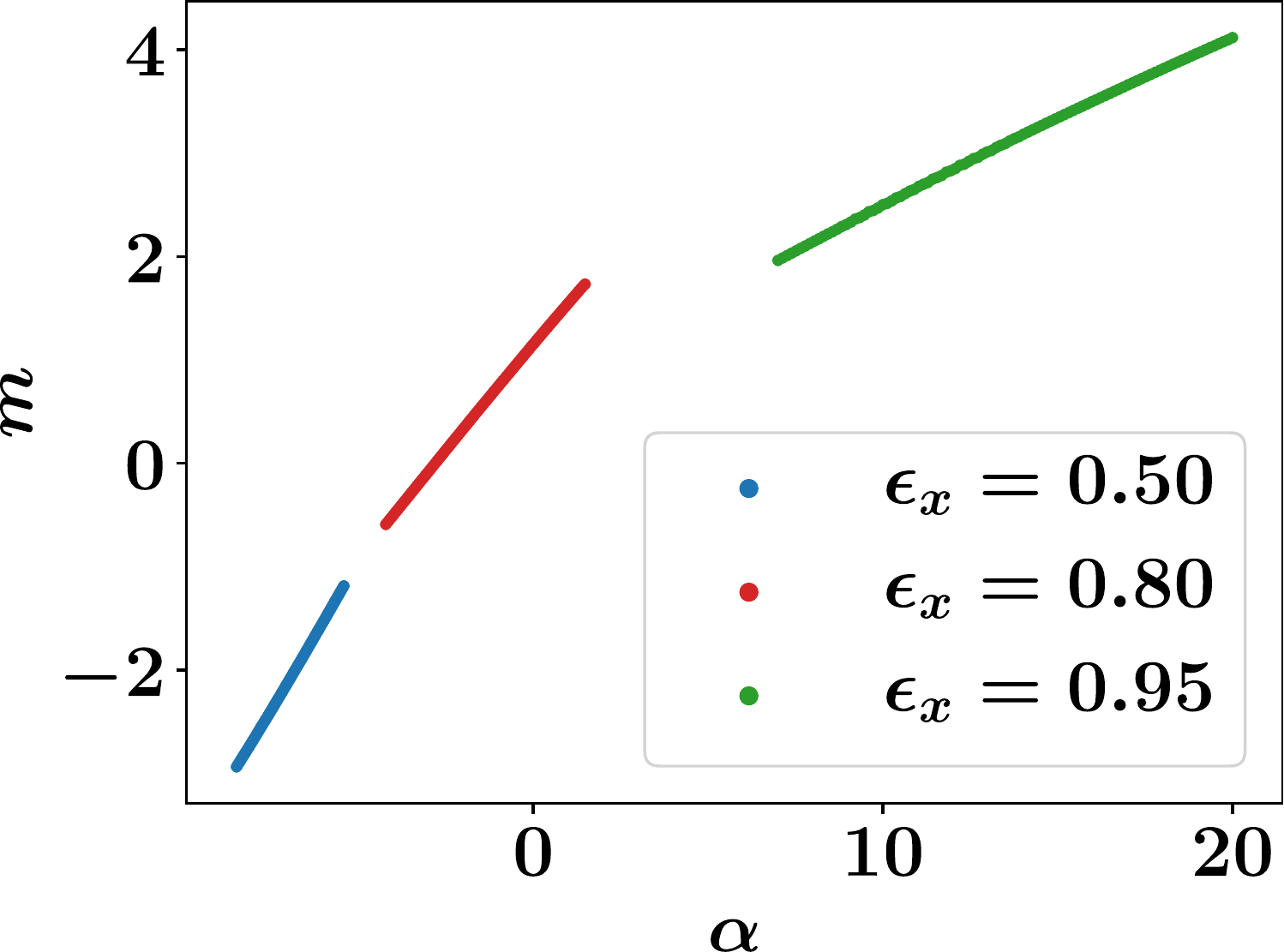}
\caption{Normalized moment $m(\alpha)$ of a crease with a characteristic size $s_0=9.2 \times 10^{-3}$ and a rest angle $\Psi_0$ ranging from 36\textdegree\,to 92\textdegree for different fixed strains $\epsilon_x$.}
\label{fig:moment_micro_macro2}
\end{center}
\end{figure}

Another relevant observation is the quasi-affine relationship between the stress $m$ and the load $\alpha$ when the strain $\epsilon_x$ is maintained constant. As a consequence, during the relaxation of the force both the easy to record force $F$ and the local stress at the crease follow the same evolution. While our following study on the temporal evolution of the crease only focuses on $F$, our models and conclusions also apply to the local mechanics (see Sec.~\ref{sec:Relaxation}).

The present pre-strained elastica model was shown to be relevant for studying the mechanical response of a fold in the elastic regime, namely as long as deformations remain small enough or when the crease was annealed beforehand~\cite{Jules2019}. Otherwise, the plastic behavior of the material within the crease comes into play. In the following, we postulate that the material's plasticity consists in modifying the absolute reference state of the fold through the intrinsic parameters $\Psi_0$ and $S_0$. This hypothesis allows us to extend our model beyond the purely elastic behavior.

However, the dominant plastic contribution comes from variations of the rest angle $\Psi_0$ (see Fig.~\ref{fig:Mapping}) which yields a powerful tool, the aforementioned mapping, that allows us to predict the full state of the fold and provide local information on the system at any time during deformation using the macroscopic observables $l$ and $F$. To test this idea in the plastic regime, we systematically compare in Section~\ref{sec:Plasticity} the prediction of the rest angle to experimental observations of $\Psi_0$.

\section{Plastic response of a crease}
\label{sec:Plasticity}

For the current experiments, we used rectangular mylar (PET) sheets of length 159~mm, width 30~mm, and thickness 500~$\mu$m. The sheet was manually pre-creased at its half-length and put under a heavy weight for 30~minutes. Then, the fold was freely let to relax for 10~minutes. Before performing the experiment, we took a high-resolution photo of the free-standing fold to measure its characteristic size $S_0^{init}$ and rest angle $\Psi_0^{init}$. These parameters were extracted by interpolating the shape of the fold using Eq.~(\ref{eqn:ref_state}). Then, a simple stretching test probed its mechanical response, as presented in Fig.~\ref{fig:SchemaMontage} using two different protocols.

In a first experiment (Protocol A), we clamped the fold in a compressed state ($F<0$) and stretched it at a speed of $50~\mathrm{mm.s}^{-1}$ until the elastic limit was crossed, and a given target force was reached. Here, we chose 4N as a maximum load to make sure the material was stressed well above its plastic threshold. Then, the fold was unclamped, and a photo of the final state was taken immediately to extract the rest angle $\Psi_0^{final}$ and the characteristic size $S_0^{final}$. In a second experiment (Protocol B), we followed the same procedure except that the fold was compressed back to a target negative force, in our case $-1.5$N, before unclamping it and measuring again $\Psi_0^{final}$ and $S^{final}$. For the 500~$\mu$m mylar sheets we used, we found $s_0^{init} \simeq 9.2 \times 10^{-3}$ and $s_0^{final} \lesssim 1.1 \times 10^{-2}$ amounting to a 20\% variation of the dimensionless characteristic size throughout the whole experiment. Fig.~\ref{fig:Mapping}(b) shows that for a fold prepared with $s_0 \lesssim10^{-2}$, the main effect on the load-displacement curve comes from variations of the rest angle $\Psi_0$. Guided by this result, we neglect in the following the measured variations of the characteristic size and assume $s_0 = s_0^{init}=s_0^{final}\simeq 10^{-2}$.

The comparison between the raw output of the experiments, {\it i.e.} the load-displacement curves, and the theoretical results of the pre-strained elastica requires a normalization factor proportional to the bending rigidity of the material, $B$. For that purpose, we use the experimental results within the elastic regime, which corresponds to an applied force $F \lesssim 0$. Then $B$ is set as a fitting parameter for the load-displacement curve while the internal parameters of the fold are given by $\Psi_{0}^{init}$ and $s_{0}$. The experimental load-deformation curve is well-reproduced by the elastic model with $B= 44.2 mJ$, which corresponds to a young modulus $E \approx 3.5$ GPa and a Poisson ratio $\nu=0.4$, consistent with tabulated values. Using the fold as a 'bendometer' for thin sheets is indeed very accurate compared to other flexural tests. For example, as opposed to an unfolded sheet submitted to the same experimental test, the advantage of using the mechanical response of a fold for measuring $B$ is that it does not experience a buckling threshold and allows for a broader range of accessible deformations.

\begin{widetext}

\begin{figure}[htb]
\begin{center}
\includegraphics[width=0.7\linewidth]{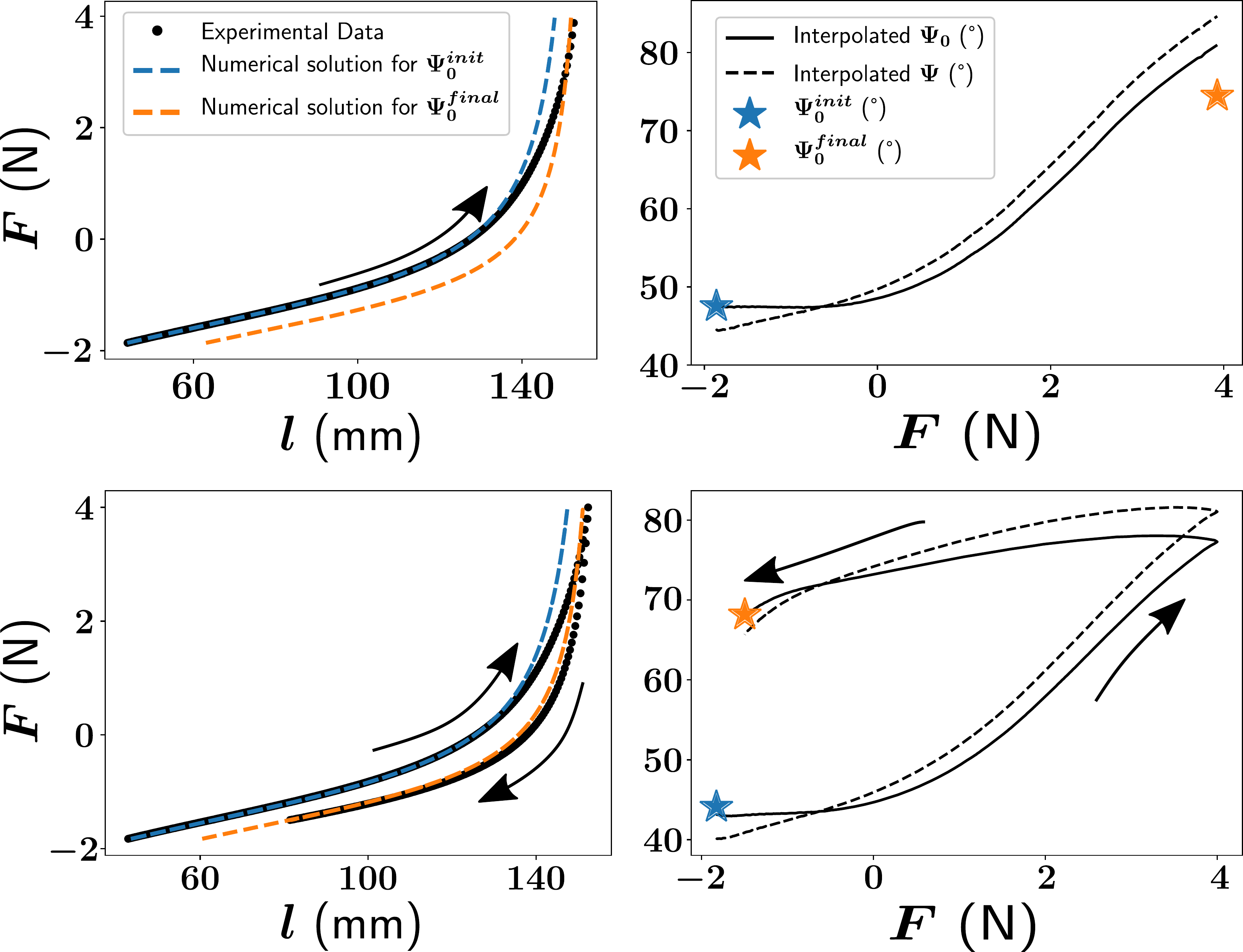}
\caption{Mechanical response from a fold made from a mylar sheet using two different protocols (see text). The top (bottom) row shows the experimental results of protocol A (B). Arrows in figures follow the chronology of the loading test. Black circles in the first column show the corresponding experimental force-displacement data. The second column shows the mapping of the folded state throughout the whole mechanical testing of (solid curve) the rest angle $\Psi_0$ in the reference state and (dashed curve) the actual angle of the crease $\Psi$ in the loaded configuration. $\Psi_{0}^{init}$ and $\Psi_{0}^{final}$ are respectively the measured rest angles of the corresponding freely standing folds before and after the mechanical test. We use these angle values in the left column to highlight deviations of the experimental mechanical response form a pure elastic behavior (orange and blue dashed curves). Finally, the legends are common to both rows.}
\label{fig:exp_mapping}
\end{center}
\end{figure}

\end{widetext}

During the whole loading test, we assume that our system is instantaneously in a quasi-static elastic equilibrium state. As a result, the deformation of the fold always follows Eqs.~(\ref{eqn:PSE},\ref{eqn:BC}). With these considerations, the plasticity of the system only translates into corrections of the rest angle $\Psi_0$ (recall that the characteristic length $s_0$ is kept constant). After appropriate normalization of the applied force using the measured value of $B$, we interpolate the crease rest angle $\Psi_0$ for each experimental situation using the mapping of the normalized ($\epsilon_x$,$\alpha$) phase space shown in Fig.~\ref{fig:Mapping}(a). The results are shown in Fig.~\ref{fig:exp_mapping} for each experimental protocol. Moreover, the actual angle $\Psi$ of the crease in the deformed configuration is a direct output of such mapping.

The interpolation of the experimental mechanical response using the mapping procedure shows two different regimes (see Fig.~\ref{fig:exp_mapping}). At small deformations where the fold responds elastically, the rest angle of the fold $\Psi_0$ is constant ($\approx\Psi_0^{init}$) as expected~\cite{Jules2019}. When the deformation is large enough, local stresses within the creased region exceed the yield stress inducing a plastic response of the material. This behavior translates in the reference configuration of the crease through a variation of the rest angle $\Psi_0$, which increases with $\epsilon_x$. In protocol~A, the measured final rest angle $\Psi_0^{final}$ differs notably from predictions which, a posteriori, is an expected result. Indeed, in addition to plastic behavior, high stresses induce a relaxation process of the crease whose amplitude is proportional to the imposed load~\cite{Thiria2011}. The few seconds between unclamping the fold and image capture are enough to change the rest angle significantly. Protocol~B addresses this issue by bringing back the fold in a compressed state where local stresses within the creased region are below the yield stress. In this case, our interpolation procedure correctly recovers the final rest angle of the crease $\Psi_0^{final}$.

Recall that the predictions of both instantaneous intrinsic parameters of the fold and the bending rigidity of the sheet result from only the interpolation of the load-displacement curve and the initial rest angle $\Psi_0^{init}$. As a result, the agreement between experimental and predictions is comforting regarding the validity of the prestrained elastica model. For both experiments, the rest angle constantly evolves above the plasticity threshold, even when the crease is brought back to low-stress configurations. When the plasticity threshold is crossed, the viscous-like behavior of the material responsible for the relaxation of the fold~ \cite{Thiria2011} creates inertia in the system and makes it difficult to reach a constant rest angle over an amount of time comparable to that of the experiment. In Section~\ref{sec:Relaxation}, we focus on the complete temporal evolution of the fold. We will, in particular, use the property of the system displayed in Fig.~\ref{fig:moment_micro_macro2} that allows us to only focus on macroscopically measured quantities, while the mapping will make the link to the internal state parameters of the fold.

\section{Aging properties of a crease}
\label{sec:Relaxation}

Previous experiments on a single fold reported temporal evolution of their shapes characterized by a simple logarithmic aging law. This behavior was observed in freely standing folds~\cite{Thiria2011,Benusiglio2012} and shown to persist when the fold underwent mechanical sollicitation~\cite{Thiria2011}. Recent studies on crumpled polymeric sheets~\cite{Lahini2017, Bruggen2019} witnessed similar relaxation phenomena. By analogy with glassy systems~\cite{Bouchaud1992, Kolvin2012, Bertin2003}, this behavior was modeled by assuming a specific distribution of microscopic timescales that produces a logarithmic temporal response of the macroscopic observable~\cite{Amir2012}. For crumpled sheets, both crease network~\cite{Gottesman2018} and friction~\cite{Mellado2011} induced by self-contact of the different parts of the sheet participate in the statistical distribution of timescales, on top of those present in the material itself. To discriminate the impact of each contribution, we concentrate in the following on the temporal behavior of a single fold.

Using the experimental setup of Fig.~\ref{fig:SchemaMontage} again, we perform relaxation experiments under imposed global strain $\epsilon_x$. Here, the displacement $l$ is fixed at values above the elastic range, while the load $F$ is recorded. The main difficulty with such experiments lies in their reproducibility, as the mechanical response of this seemingly simple system is history-dependent while the study of aging properties requires using the same fold for each series of experiments. To address this problem, we first laid the fold after each mechanical testings under a heavy weight for 30 minutes. The procedure aims to `reset' the initial state of the crease. Then, we leave the fold freely relaxing for a small timelapse. This transition alleviates the temporal dependence at the beginning of each experiment~\cite{Thiria2011}.

\begin{figure}[htb]
\begin{center}
\includegraphics[width=0.9\linewidth]{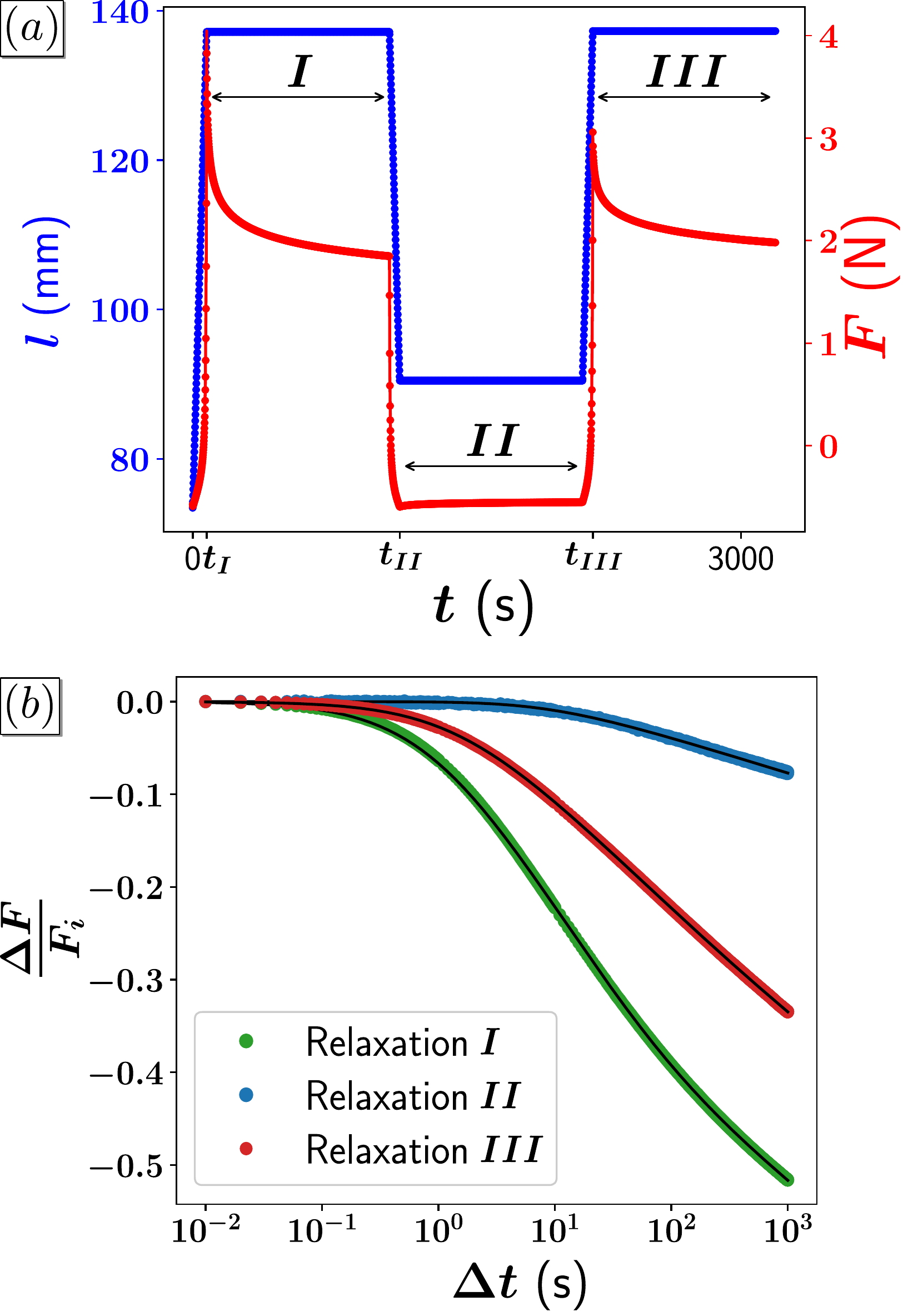}
\caption{(a) Protocol for studying relaxation of a fold: the displacement $l(t)$ is prescribed while the instantaneous response $F(t)$ is recorded. The chosen protocol defines three relaxation phases starting at $t=t_i$ corresponding to applied forces $F_i$ ($i=I,II,III$). (b) Relaxation of the force during the three relaxation phases compared to the double-logarithmic behavior given by Eq.~(\ref{eq:relaxfit}) (solid lines). Here, $\Delta t= t-t_i$ and $\Delta F = F(t)-F_i$.}
\label{fig:ProtocolRelax}
\end{center}
\end{figure}

A typical experiment is divided into at least three relaxation phases, $I$, $II$ and $III$ (see Fig.~\ref{fig:ProtocolRelax}(a)). Each relaxation experiment follows the same protocol: we monotonically vary the gap distance $l(t)$ until the force reaches an extremal value $F_i$ at a time $t_i$ ($i=I,II,III$). Then, the gap distance is kept fixed, and the temporal evolution of the force $F(t)$ is recorded. For each series of experiments on the same fold, the prescribed $l(t)$ for relaxation phases $I$ and $II$ is kept identical, while phase $III$ may change between different experimental runs by varying the extremal force $F_{III}>0$. This protocol serves two goals: to check the reproducibility of the results and to prepare the system in the same state before the last relaxation phase.

The aging behavior of the fold has been characterised by a single logarithmic time evolution~\cite{Thiria2011}. However, such description fails to capture the complete, nonmonotonic temporal evolution encountered in our experiment. To this purpose, we interpolate the force signals for the different relaxation phases using a double-logarithmic function given by~\cite{Lahini2017}
\begin{equation}
\frac{\Delta F}{F_i} = A_1 \log\left(1+\frac{\Delta t}{\tau_1}\right)+A_2\log\left(1+\frac{\Delta t}{\tau_2}\right),
\label{eq:relaxfit}
\end{equation}
where $\Delta t=t-t_i$ and $\Delta F=F(t)-F_i$ with $F_i=F(t_i)$. Eq.~(\ref{eq:relaxfit}) involves two relaxation rates $(A_1,A_2)$ and two time scales $\tau_1 < \tau_2$. Fig.~\ref{fig:ProtocolRelax}(b) shows that the double-logarithmic interpolation describes well experimental results in the different relaxation phases. In phase~I, we found $A_1 = -0.091 \pm 0.002$, $A_2 = 0.048 \pm 0.002$, $\tau_1 = 0.9 \pm 0.07$~s and $\tau_2 = 83 \pm 9$~s. In phase~$II$, the force signal exhibits a simple logarithmic decay with $A_1 = -0.016 \pm 0.0007$, $A_2 = 0$ and $\tau_1 = 4.4 \pm 1.6$~s. For all experiments, the fitting parameters in phase $I$ and $II$ are found to be consistently constant.  This two-phase preparation thus achieves both objectives: the sample is left in a compressed state before phase~III with reproducible response and a controlled short-term history. Since we use the same sample for all the experiments, our `reset' procedure of the initial state also succeeds in limiting the impact of the long-time history.

\begin{figure}[htb]
\begin{center}
\includegraphics[width=0.8\linewidth]{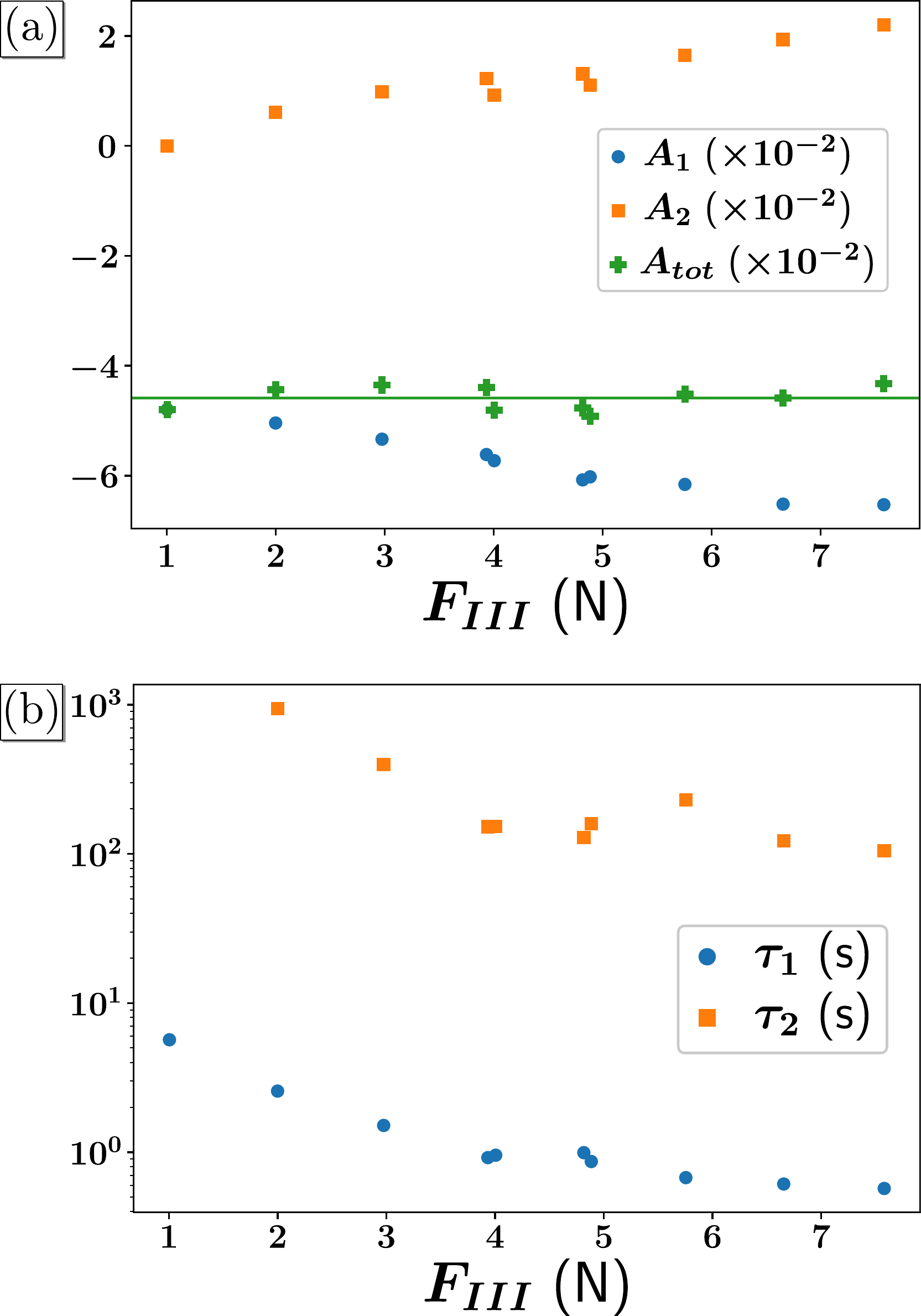}
\caption{Influence of the extremal force $F_{III}$ on the parameters of the double-logarithmic interpolation of phase~III relaxation for the same fold in different experiments. a) Evolution with $F_{III}$ of the relaxation rates $A_1$, $A_2$ and their sum $A_{tot} = A_1 + A_2$. The solid line shows the mean $<A_{tot}>=- 0.046$. b) Semi-Log plot of the time scales $\tau_1$ and $\tau_2$ as function of $F_{III}$. }
\label{fig:EffectForce}
\end{center}
\end{figure}

In phase~$III$, we varied the value of the extremal force $F_{III}$ and looked at its effect on the relaxation of the fold. The results are shown in Fig.~\ref{fig:EffectForce}. For every experiment we found $\tau_1 \ll \tau_2$, which is consistent with relaxation in phase~$I$ and indicates that Eq.~(\ref{eq:relaxfit}) describes effectively two separate phenomena. Both $\tau_1$ and $\tau_2$ are found to decrease with $F_{III}$, while the absolute amplitude of each logarithmic term increases with $F_{III}$. These observations point to a repartition of multiple time scales shortened by the increase of the local stress. Surprisingly, the long-time relaxation rate $A_{tot} = A_1+A_2\approx -0.046$ is constant regardless of the imposed macroscopic stress. This result is consistent with the long time relaxation behavior reported in~\cite{Thiria2011}.

\begin{figure}[htb]
\begin{center}
\includegraphics[width=0.9\linewidth]{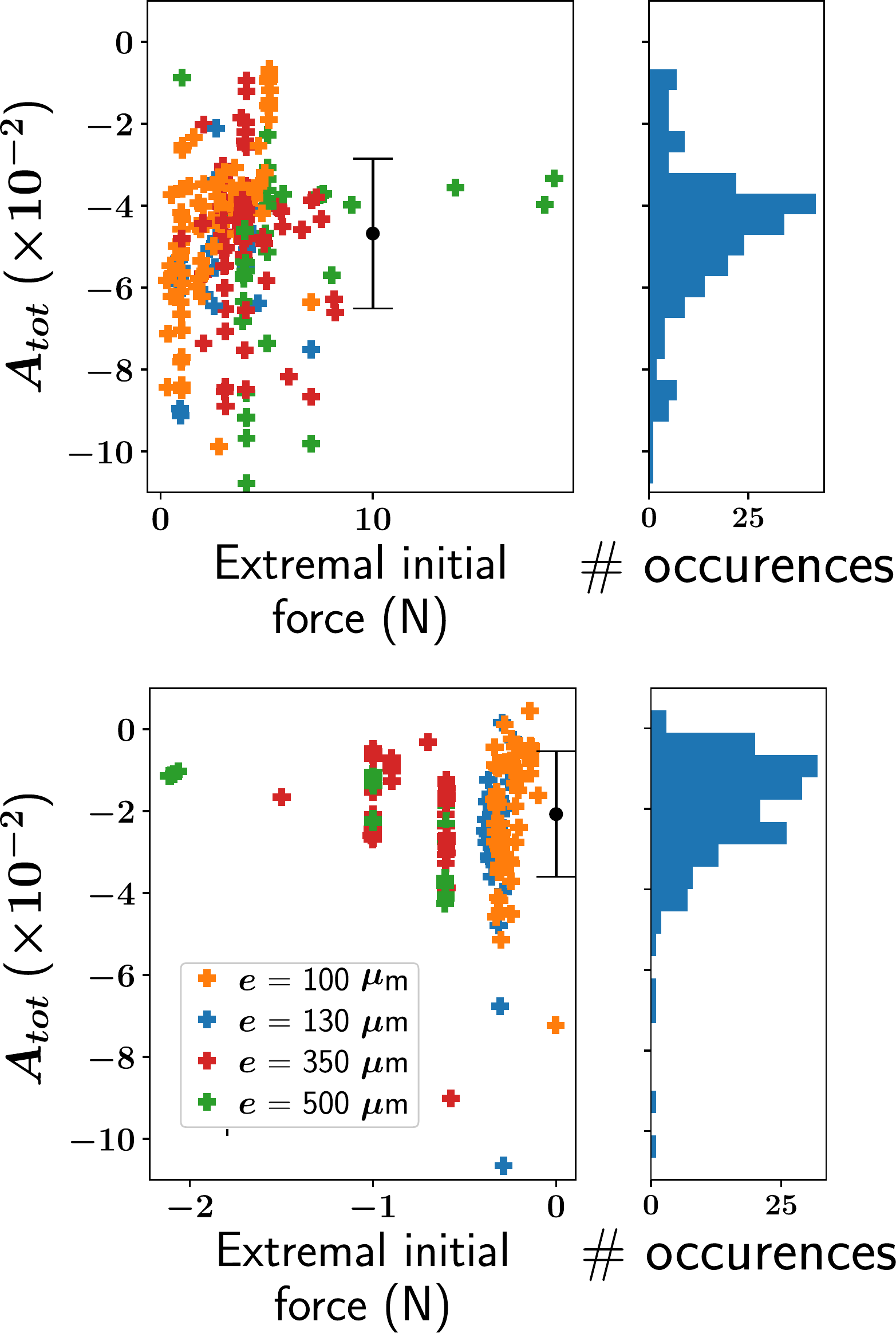}
\caption{Effects of the experimental conditions on the relaxation rate $A_{tot}$ (see text). Each point in the left column corresponds to different experimental conditions. For convenience, the data is shown as a function of the extremal force at the beginning of the relaxation process, and colors label different material thicknesses $e$. The mean (black point) and standard deviation of all data points are also shown. The top (bottom) row corresponds to folds under tension (compression). The right column shows the histogram of $A_{tot}$.}
\label{fig:AmplitudeAllRelax}
\end{center}
\end{figure}

One notices that the relaxation rates $A_{tot}$ are very similar for phases $I$ ($A_{tot}\approx-0.043$) and $III$ ($A_{tot}\approx -0.046$) showing that while the preparation of the fold is very different, the qualitative behavior of the relaxation is robust. To test this feature, we performed 405 relaxation experiments by modifying the experimental system in several ways. The changes include varying the fabrication process of the crease, the number of relaxation cycles in a single experiment, the value of the extremal force $F_i$ for each relaxation, the timespan of the relaxation, the dimensions of the fold (length, width, and thickness) and the ambient temperature (from 5\textdegree C\, to 45\textdegree C) using a controlled bath. The corresponding amplitudes for all relaxation rates are gathered in Fig.~\ref{fig:AmplitudeAllRelax}. As expected, the normalized relaxation rate at long times $A_{tot}$ varies depending on the preparation and on the experimental conditions. However, some general trends are common to all relaxations. For instance, the amplitude $A_{tot}$ is always negative and remains of the same order of magnitude for all experiments, with typically $A_{tot} = -0.047 \pm 0.018$ for stretched folds and $A_{tot} = -0.021 \pm 0.015$ for compressed ones. Therefore, if one considers $A_{tot}$ as the main characteristic relaxation rate, these results are consistent with previous work~\cite{Lahini2017,Thiria2011,Bruggen2019}.

\begin{figure}[htb]
\begin{center}
\includegraphics[width=0.9\linewidth]{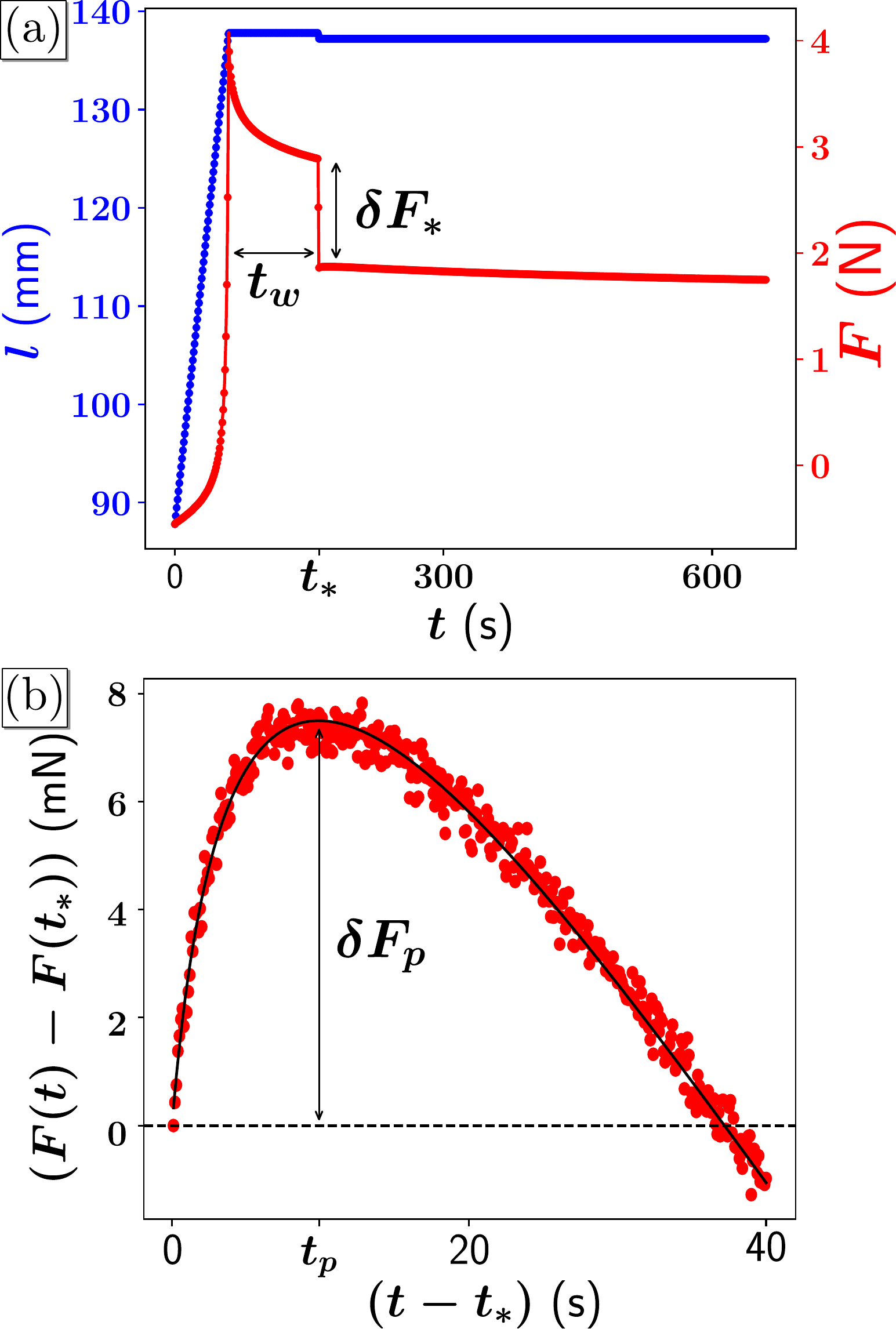}
\caption{(a) Modified protocol during the relaxation phase~$III$ to obtain non-monotonic relaxation. After a waiting time $t_w$ (here $t_w = 100~\text{s}$), a jump in displacement is imposed to lower the force by an amount $\delta F_{*}$ (here $\delta F_*=1~\text{N}$). (b) Non-monotonous relaxation of the force signal for $t>t_*$. The relative force $F(t)-F(t_*)$ increases up to a maximum $\delta F_p$ at a peak time $t-t_*=t_p$ before decreasing again. The black line is a fit by the double-logarithm in Eq.~(\ref{eq:relaxfit}) with $A_1 = 0.008$, $A_2 = -0.11$, $\tau_1 = 1.9$ s and $\tau_2 = 156$ s.}
\label{fig:ProtocolRelaxNM}
\end{center}
\end{figure}

The double-logarithmic aging behavior hints at the presence of multiple time-scales in a folded structure. This feature was shown in crumpled sheets~\cite{Lahini2017} using a specific experiment devised to create non-monotonic relaxation: a crumpled mylar sheet is put under heavy weight for a given duration, before slightly easing the compression and measuring the relaxation of the external load. Drawing inspiration from this work, we modified our initial protocol by adding a new step during phase~$III$ of the mechanical test. Fig.~\ref{fig:ProtocolRelaxNM}(a) shows a fold that is let to relax during a waiting time $t_w$ before instantaneously decreasing the imposed displacement by $\delta l_{*}$ which in turn lowers the force by an amount $\delta F_{*}$. Fig.~\ref{fig:ProtocolRelaxNM}(b) shows that the following relaxation indeed displays a non-monotonic evolution of the force, similar to the one found in crumpled sheets~\cite{Lahini2017}.

\begin{figure}[htb]
\begin{center}
\includegraphics[width=0.8\linewidth]{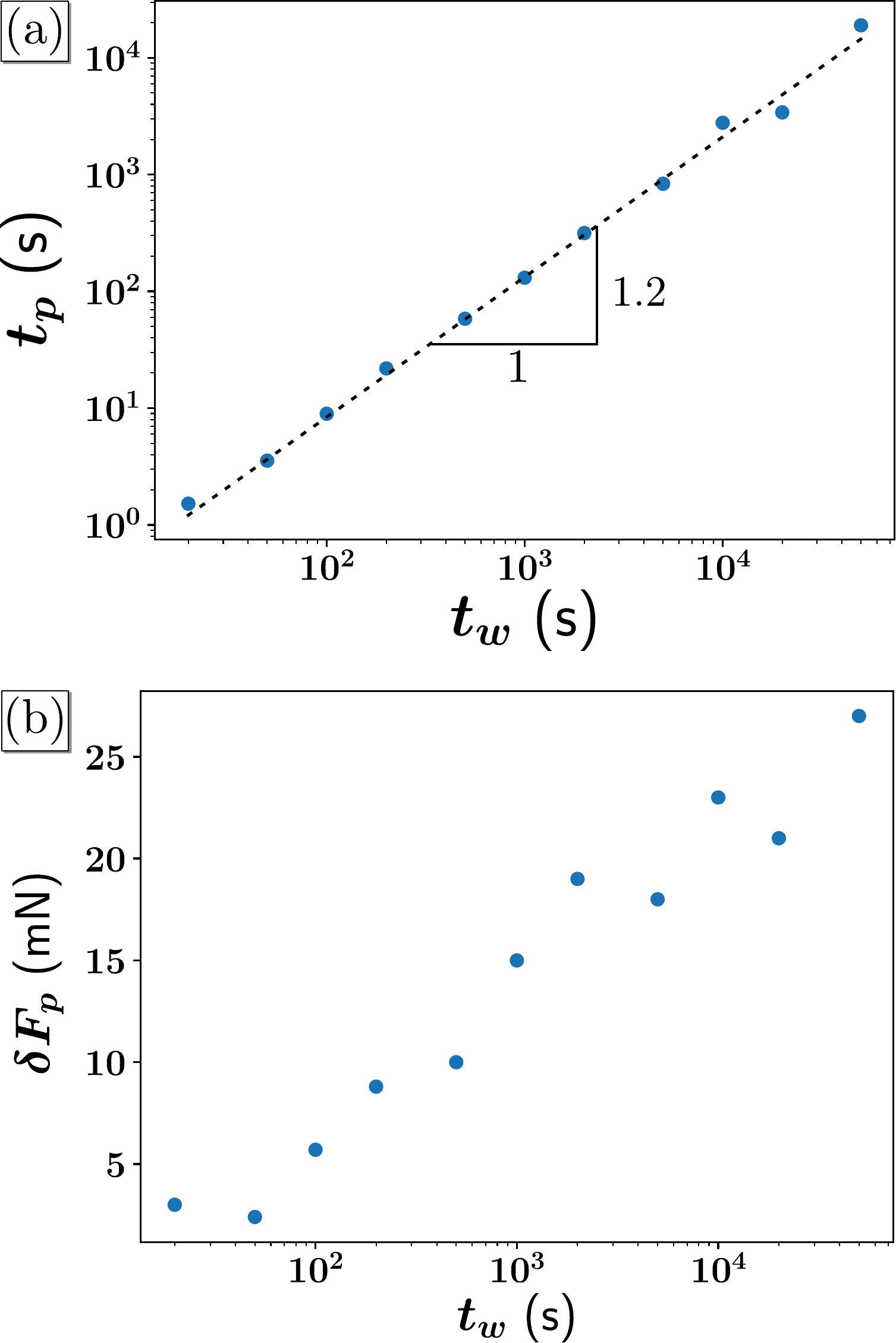}
\caption{(a) The peak time $t_p$ as functions of the waiting time $t_w$ for relaxation experiment shown in Fig.~\ref{fig:ProtocolRelaxNM}. The dashed line shows a scaling behavior $t_p \propto t_w^{1.2}$. (b) Amplitude of the maximum relative force $\delta F_p$ as function of the waiting time. }
\label{fig:AllRepeak}
\end{center}
\end{figure}

A systematic study of the effect of the waiting time $t_w$ on the relaxation is shown in Fig.~\ref{fig:AllRepeak}. As expected, the time $t_p$ at the peak of the relaxation increases with $t_w$. However, the observed weakly nonlinear scaling $t_p \propto t_w^{1.2}$ does not coincide with the linear behavior $t_p \propto t_w$ observed in crumpled sheets~\cite{Lahini2017} nor with a naive dimensional analysis. This nonlinear behavior hints for the existence of a characteristic time scale whose origin is still unclear. However, we expect that this relationship depends on the specific rheology of the material, which induces complex long term memory effects and thus a distribution of time scales. Interestingly, Fig.~\ref{fig:AllRepeak}(b) shows that the amplitude of the force anomaly grows with the logarithm of the waiting time, pointing towards a collection of activated mechanisms.

\section{Conclusion}
\label{sec:Discussion}

Our study has heavily relied on experiments to identify and thoroughly characterize the two dominant sources of irreversibility that arise during the mechanical solicitation of material creased sheets: plasticity and slow relaxation. We have shown that an elastic model introduced earlier for the fold can be refined to capture the plastic flow of the system fully: when the crease is localized, this flow only amounts to a change the crease reference angle, while the rest of the system remains elastic. This approach provides a powerful relationship between macroscopic mechanical observables, that can easily be measured, and the microscopic state of the crease, in particular its rest angle. The relevance of this approach is emphasized by the demonstrated shallowness of the elastic regime in the fold, and by the fact that it is expected to hold for a wide range of materials including polymers and metal. Furthermore, within our testing configuration and due to the strong lever effect involved, the fold acts as a 'bendometer', as the formalism we developed allows for precise measurement of the bending modulus of the underlying material.

Despite complex memory effects, we drew inspiration from glassy systems \cite{Bertin2003} to rationalize the temporal behavior of the observed mechanical response, invoking a distribution of time-scales within the material to explain the slow relaxation and contingent non-monotonicity of the constraints for a given deformation path. In this respect, these results are specific to materials with non-trivial rheology, in particular glassy materials~\cite{Amir2012}. However, the qualitative agreement between the temporal responses a single fold we observe and that of a crumpled polymeric sheet~\cite{Lahini2017} questions the role of the collective phenomena in the latter results. The crease itself magnifies the material response and already introduces at the individual level the complexity the authors observe in the extended system. Still, the high number of creases in crumpled sheets might smoothen the mechanical evolution of the system, leading to less memory dependent single logarithmic relaxations.

Finally, our study demonstrates the predictive power of a continuous description in the single crease problem, as embodied by our pre-strained elastica model, and beyond. Indeed, it can be generalized to more complex, extended patterns to infer very strong constraints on their equilibrium configuration, and to gain insight into their mechanical response, including plasticity and aging. Our study thus lays the foundations of a universal approach to the mechanics of a class of systems encompassing structured origamis and crumpled material sheets.

\section*{Acknowledgements}

We thank Yoav Lahini for interesting discussions.

\bibliography{Biblio}

\end{document}